%% file: Paper.tex
\documentclass[11pt,letterpaper]{article}

\usepackage{mathrsfs}           
\usepackage{vmargin,fancyhdr}   
\usepackage{algorithm}
\usepackage{algorithmic}

\usepackage{amsmath,amssymb}    
\usepackage{verbatim}           
\usepackage{xspace}             
\usepackage{graphicx,float}     
\usepackage{ifthen,calc}        
\usepackage{textcomp}           
\usepackage{fancybox}           
\usepackage{hhline}             
\usepackage{float}              

\usepackage[rflt]{floatflt}     
\usepackage[small,compact]{titlesec}
\usepackage{setspace}
\usepackage{subfigure}

\usepackage{color}
\definecolor{ForestGreen}{rgb}{0.1333,0.5451,0.1333}
\usepackage[letterpaper,dvips,
            colorlinks,linkcolor=ForestGreen,citecolor=ForestGreen,
            backref,
            bookmarks,bookmarksopen,bookmarksnumbered]
           {hyperref}
\usepackage[dvips]{thumbpdf}

\setpapersize{USletter}
\setmarginsrb{.75in}{.5in}        
             {.75in}{.5in}        
             {.25in}{.25in}     
             {.25in}{.5in}      
\newcommand{\showccc}[0]{0}
\newcommand{\ccc}[2][nothing]{
  \ifthenelse{\showccc=0}{}{
    \ensuremath{^{\Lsh\Rsh}}\marginpar{\raggedright\tiny\textsf{%
        \ifthenelse{\equal{#1}{nothing}}{}{\textbf{#1}\\}#2}}}}

\newcounter{hours}\newcounter{minutes}
\newcommand{\hhmm}{%
  \setcounter{hours}{\time/60}%
  \setcounter{minutes}{\time-\value{hours}*60}%
  \ifthenelse{\value{hours}<10}{0}{}\thehours:%
  \ifthenelse{\value{minutes}<10}{0}{}\theminutes}
\ifthenelse{\showccc=0}{\rhead{}}{\rhead{\today \ [\hhmm]}}
\newtheorem{theorem}{Theorem}[section]

\newtheorem{definition}[theorem]{Definition}

\newtheorem{lemma}[theorem]{Lemma}

\newcommand{\Proof}[0]{\smallskip\noindent\textit{\textbf{Proof}}\quad}

\newcommand{\QED}[0]{\hfill\ensuremath{\blacksquare}\medspace}

\newcommand{\R}[0]{\ensuremath{\mathbb{R}}}

\floatstyle{ruled}
\newfloat{algo}{tbp}{lop}
\floatname{algo}{Algorithm}

\newcommand{\norm}[1]{\left\Vert#1\right\Vert}

\usepackage{float}
\floatstyle{plain}\newfloat{myfig}{t}{figs}[section]
\floatname{myfig}{\textsc{Figure}}
\floatstyle{plain}\newfloat{myalg}{H}{algs}[section]
\floatname{myalg}{}
\begin{document} 

\title{Approaching optimality for solving SDD linear systems\thanks{Partially supported by the National Science Foundation under grant number CCF-0635257.}
}
\author{
  Ioannis Koutis\thanks{Partially supported by Microsoft Research through the Center for Computational Thinking at CMU}
  \quad
  Gary L.\ Miller
  \quad
  Richard Peng\thanks{Partially supported by Natural Sciences and Engineering Research Council of Canada (NSERC) under grant number M-377343-2009.}
  \quad\\
  Computer Science Department\\
  Carnegie Mellon University \\
  \texttt{\{ioannis.koutis,glmiller,yangp\}@cs.cmu.edu}
}
\maketitle

\input{abstract}
\input{intro}
\input{background}

\input{contributions}

\input{sample}

\input{incrementalsparsify}

\input{solver}

\input{extensions}






\begin{spacing}{0.7}
  \begin{small}
    \bibliographystyle{alpha}
    \input{Paper.bbl}

  \end{small}
\end{spacing}

\newpage

\input{rpcheb}

\end{document}

%% file: abstract.tex
\begin{abstract}
\textnormal {We present an algorithm that on input of an $n$-vertex $m$-edge weighted graph $G$ and a value $k$, produces an {\em incremental sparsifier} $\hat{G}$ with $n-1 + m/k$ edges, such that the condition number of $G$ with $\hat{G}$ is bounded above by $\tilde{O}(k\log^2 n) \footnote{We use the $\tilde{O}()$ notation to hide a factor of at most $(\log\log n)^4$}$, with probability $1-p$. The algorithm runs in time
$$\tilde{O}((m \log{n} + n\log^2{n})\log(1/p)).$$
As a result, we obtain an algorithm that on input of an $n\times n$ symmetric diagonally dominant matrix $A$ with $m$ non-zero entries and a vector $b$, computes a vector ${x}$ satisfying $||{x}-A^{+}b||_A<\epsilon ||A^{+}b||_A $, in expected time
$$\tilde{O}(m\log^2{n}\log(1/\epsilon)).$$
The solver is based on repeated applications of the incremental sparsifier that produces a chain of graphs which is then used as input to a recursive preconditioned Chebyshev iteration.}
\end{abstract}

%% file: intro.tex
\section{Introduction}

   Fast algorithms for solving  linear systems and the related problem of finding a few fundamental eigenvectors is
   possibly one of the most important problems in algorithm design.  It has motivated work on fast matrix multiplication    methods, graph separators,  and more recently graph sparsifiers.  For most applications the matrix is sparse, and
   thus one would like algorithms whose run time is efficient in terms of the number of non-zero entries of the matrix.
   Little is known about how to efficiently solve general sparse systems, $Ax =b$.  But substantial progress has been
   made in the case of symmetric and diagonally dominant (SDD) systems, where $A_{ii} \geq \sum_{j \not= i} |A_{ij}|$.    In a seminal work, Spielman and Teng showed that SDD systems can be solved in nearly-linear time \cite{SpielmanTeng04,EEST05,SpielmanTeng08c}.

   Recent research, largely motivated by the Spielman and Teng solver (ST-solver), demonstrates the power of SDD solvers as an algorithmic primitive. The ST-solver is the key subroutine of the fastest known algorithms for a multitude of problems that include: (i)  Computing the first non-trivial (Fiedler) eigenvector of the graph, or more generally the first few eigenvectors, with well known applications to the sparsest-cut problem \cite{Fiedler73,SpielmanTeng96,chung1}; (ii) Generating spectral sparsifiers that also act as cut-preserving sparsifiers \cite{SpielmanSrivastava08}; (iii) Solving linear systems derived from elliptic finite elements discretizations of a significant class of partial differential equations \cite{BHV04}. (iv) Generalized lossy flow problems \cite{SpielmanDaitch08}; (v) Generating random spanning trees \cite{kernerFaster09}; and (vi)  Several optimization problems in computer  vision \cite{KoutisMT09,KMST-TR-09} and graphics \cite{mccannreal08,pushkarHarmonic07};  A more thorough discussion of applications of the solver can be found in \cite{Spielman10Survey, Teng10Survey}.

   The ST-solver is an {\em iterative} algorithm that produces a sequence of approximate solutions converging to the actual solution of the input system $Ax=b$.  The performance of iterative methods is commonly measured in terms of the time required to reduce an appropriately defined approximation error by a constant factor.  Even including recent improvements on some of its components, the time complexity of the ST-solver is at least $O(m\log^{15} n)$. The large exponent in the logarithm is indicative of the fact that the algorithm is quite complicated and lacks practicality.
   The design of a faster and simpler solver is a challenging open question.

   In this paper we present a conceptually simple and possibly practical iterative solver that runs in time $\tilde{O}(m \log^2 n)$. Its main ingredient is a new {\em incremental} graph sparsification algorithm, which is of independent interest. The paper is organized as follows. In Section \ref{sec:background} we review basic notions and we introduce notation. In Section \ref{sec:history} we discuss the development of SDD solvers, the algorithmic questions it motivates, and the progress on them, with an emphasis on the graph sparsification problem. In Section \ref{sec:techniques} we present a high level description of our approach and discuss implications of our solver for the graph sparsification problem.   The incremental sparsifier is presented and analyzed in Sections \ref{sec:sample} and \ref{sec:incrementalsparsify}. In Section \ref{sec:solver} we explain how it can be used to construct the solver. Finally, in the Appendix we give pseudocode for the complete solver. 


%% file: background.tex
\section{Preliminaries} \label{sec:background}

In this Section we briefly recall background facts about Laplacians of weighted graphs. For more details, we refer the reader to \cite{RoyleGodsil} and \cite{Boman03support}. Throughout the paper, we discuss connected graphs with positive edge weights. We use $n$ and $m$ to denote $|V|$ and $|E|$.

A symmetric matrix $A$ is positive semi-definite if for any vector $x$, $x^T Ax \geq 0$. For such semi-definite matrices $A$, we can also define the $A$-norm of a vector $x$ by
$$||x||_A^2=x^T A x.$$

Fix an arbitrary numbering of the vertices and edges of a graph $G$. Let $w_{i,j}$ denote the weight of the edge $(i,j)$. The Laplacian $L_G$ of $G$ is the matrix defined by: (i) $L_G(i,j) = -w_{i,j}$, (ii) $L_G(i,i) = \sum_{i\neq j} w_{i,j}$. For any vector $x$, one can check that

$$x^TL_Gx = \sum_{u, v \in E} (x_u-x_v)^2w_{uv}.$$

It follows that $L_G$ is positive semi-definite and $L_G$-norm is a valid norm.

We also define a partial order $\preceq$ on symmetric semi-definite matrices, where $A \preceq B$ if $B-A$ is positive semi-definite. This definition  is equivalent to $x^TAx \leq x^TBx$ for all $x$. We say that a graph $H$ {\em $\kappa$-approximates} a graph $G$ if
$$L_H \preceq L_G  \preceq  \kappa L_H.$$

By the definition of $\preceq$ from above, this relationship is equivalent to $x^TL_Hx \leq  x^TL_Gx  \leq \kappa x^TL_Hx$ for all vectors $x$. This implies that the {\em condition number} of the pair $(L_G,L_H)$ is upper bounded by $\kappa$. The { condition number} is an algebraically motivated notion; upper bounds on it are used to predict the convergence rate of iterative numerical algorithms.

%% file: contributions.tex
\section{Prior work on SDD solvers and related graph theoretic problems} \label{sec:history}

   Symmetric diagonally dominant systems are linear-time reducible to linear systems whose matrix is the {Laplacian} of a weighted graph via a construction known as double cover that only doubles the number of non-zero entries in the system \cite{GrMiZa95,gremban-thesis}. The one-to-one correspondence between graphs and their Laplacians allows us to focus on weighted graphs, and interchangeably use the words graph and Laplacian.


   In a ground-breaking approach, Vaidya \cite{Vaidya91} proposed the use of spectral graph-theoretic properties for the design of provably good graph {\em preconditioners}, i.e. graphs that -in some sense- approximate the input graph, but yet are somehow easier to solve. Many authors built upon the ideas of Vaidya, to develop {\em combinatorial preconditioning}, an area on the  border of numerical linear algebra and spectral graph theory \cite{BernGil-Support-02}. The work in the present paper as well as the Spielman and Teng solver is based on this approach. It is worth noting that combinatorial preconditioning is only one of the rich connections between combinatorics and linear algebra \cite{chung1,RoyleGodsil}.

   Vaidya originally proposed the construction of a preconditioner for a given graph, based on a maximum weight spanning tree of the graph and its subsequent augmentation with graph edges. This yielded the first non-trivial results, an $O((dn)^{1.75})$ time algorithm for maximum degree $d$ graphs, and an $O((dn)^{1.2})$ algorithm for maximum degree $d$ planar graphs \cite{joshi}.

   Later, Boman and Hendrickson \cite{Boman03support} made the crucial observation that the notion of {\em stretch} (see Section \ref{sec:incrementalsparsify} for a definition) is crucial for the construction of a good spanning tree preconditioner; they showed that if the non-tree edges have average stretch $s$ over a spanning tree, the spanning tree is an $O(sm)$-approximation of the graph. Armed with this observation and the low-stretch tree of Alon et al. \cite{alon95graph-theoretic}, Spielman and Teng \cite{SpielmanTeng03} presented a solver running in time $O(m^{1.31})$.

   The utility of low-stretch trees in SDD solvers motivated further research on the topic. Elkin et al. \cite{EEST05} gave an $O(m\log^2n)$ time algorithm for the computation of spanning trees with total stretch $\tilde{O}(m\log^2n)$. More recently, Abraham et. al. presented a nearly tight construction of low-stretch trees \cite{AbrahamBN08}, giving an $O(m\log n+ n\log^2n)$ time algorithm that on input a graph $G$ produces a spanning tree of total stretch $\tilde{O}(m\log n)$.  The algorithm of \cite{EEST05} is a basic component of the ST-solver. While the algorithm of \cite{AbrahamBN08} didn't improve the ST-solver, it is indispensable to our upper bound.

   The major new notion introduced by Spielman and Teng \cite{SpielmanTeng04} in their nearly-linear time algorithm was that of a {\em spectral sparsifier},  i.e. a graph with a nearly-linear number of edges that $\alpha$-approximates a given graph for a constant $\alpha$.  Before the introduction of spectral sparsifiers, Bencz{\'u}r and Karger \cite{BenczurK96} had presented an $O(m\log^3 n)$ algorithm for the construction of a {\em cut-preserving} sparsifier with $O(n \log n)$ edges. A good spectral sparsifier is a also a good cut-preserving sparsifier, but the opposite is not necessarily true.

   The ST-solver \cite{SpielmanTeng04} consists of a number of major algorithmic components. The base routine is a local    partitioning algorithm which is the main subroutine of a global nearly-linear time partitioning algorithm. The
   partitioning algorithm is used as a subroutine in the construction of the {spectral sparsifier}. Finally, the spectral sparsifier is  combined with the ${O}(m\log^2n)$ total stretch spanning trees of \cite{EEST05} to produce a $(k,O(k \log^{c} n))$ {\em ultrasparsifier}, i.e. a graph  $\hat{G}$ with $n-1+(n/k)$ edges which $O(k \log^{c} n)$-approximates the  given graph, for some $c>25$. The bottleneck in the complexity of the ST-solver lies  in the running time of the ultra-sparsification algorithm and the approximation quality of the ultrasparsifier.

   In the special case of planar graphs the ST-solver runs in time $\tilde{O}(n\log^2 n)$. An asymptotically optimal linear work algorithm for planar graphs was given in \cite{KoutisMi07}. The key observation in \cite{KoutisMi07} was that despite the fact that planar graphs don't necessarily have spanning  trees of average stretch less than $O(\log n)$, they still have $(k,c k \log k)$ ultrasparsifiers for a large enough constant $c$; they can be obtained by finding ultrasparsifiers for constant size subgraphs that contain most of the edges of the graph, and conceding the rest of the edges in the global ultrasparsifier. In addition, a more practical approach to the construction of constant-approximation preconditioners for the  case of graphs of bounded average degree was given in \cite{KoutisMiller08}. To this day, the only known   improvement for the general case was obtained by  Andersen et.al \cite{AndersenLocal06} who presented a faster and  more effective local partitioning routine that can replace the partition routine of the spectral sparsifier, improving the complexity of the solver as well.

   Significant progress has been made on the spectral graph sparsification problem. Spielman and Srivastava \cite{SpielmanSrivastava08} showed how to construct a much stronger spectral sparsifier with $O(n\log n)$ edges, by sampling edges with probabilities proportional to their effective resistance, if the graph is viewed as an electrical network.  While the algorithm is conceptually simple and attractive, its fastest known implementation still relies on the ST-solver. Leaving the envelope of nearly-linear time algorithms Batson, Spielman and Srivastava \cite{BatsonSS09} presented a polynomial time algorithm for the construction of a ``twice-Ramanujan'' spectral sparsifier with a nearly optimal linear number of edges. Finally, Kolla et al. \cite{Kolla10} gave a polynomial time algorithm for the construction of a nearly-optimal $(k,\tilde{O}(k \log n))$ ultrasparsifier.

\section{Our contribution} \label{sec:techniques}

  In an effort to design a faster sparsification algorithm, we ask: when and why the much simpler faster cut-preserving
  sparsifier of \cite{BenczurK96} fails to work as a spectral sparsifier? Perhaps the essential example is that of the
  cycle and the line graph; while the two graphs have roughly the same cuts, their condition number is $O(n)$. The
  missing edge has a stretch of $O(n)$ through the rest of the graph, and thus it has high effective resistance; the
  effective resistance-based algorithm of Spielman and Srivastava would have kept this edge. It is then natural to try
  to design a sparsification algorithm that avoids precisely to generate a graph whose ``missing'' edges have a high
  stretch over the rest of the original graph.

  This line of reasoning leads us to a conceptually simple sparsification algorithm: Find a low-stretch spanning tree
  with a total stretch of $O(m\log n)$. Scale it up by a factor of $k$ so the total stretch is $O(m \log{n} / k)$ and
  add the scaled up version to the sparsifier. Then over-sample the rest of the edges with probability proportional to
  their stretch over the scaled up tree, taking $\tilde{O}(m\log^2 n/k)$ samples. In Sections \ref{sec:sample} and
  \ref{sec:incrementalsparsify} we analyze a slight variation of this idea and we show that while it doesn't produce an
  ultrasparsifier, it produces what we call an {\em incremental sparsifier} which is a graph with $n-1+m/k$ edges that
  $\tilde{O}(k\log^2 n)$-approximates the given graph  \footnote{In the latest version of their paper
  \cite{SpielmanTeng08c}, Spielman and Teng also construct and use an incremental sparsifier, but they still use the
  term ultrasparsifier for it.}. Our proof relies on the machinery developed by Spielman and Srivastava \cite{SpielmanSrivastava08}.

  As we explain in Section \ref{sec:solver} the incremental sparsifier is all we need to design a solver that  runs in the claimed time. Precisely, we prove the following.
 \begin{theorem}
 On input an $n\times n$ symmetric diagonally dominant matrix $A$ with $m$ non-zero entries and a vector $b$,  a vector ${x}$ satisfying $||{x}-A^{+}b||_A<\epsilon ||A^{+}b||_A $, can be computed in expected time $\tilde{O}(m\log^2{n}\log(1/\epsilon)).$
\end{theorem}

\subsection{Implications for the graph sparsification problem}

The only known nearly-linear time algorithm that produces a spectral sparsifier with $O(n\log n)$ edges is due to
Spielman and Srivastava \cite{SpielmanSrivastava08} and it is based on $O(\log{n})$ calls to a SDD linear system
solver. Our solver brings the running time of the Spielman and Srivastava algorithm to $\tilde{O}(m\log^3{n})$. It is
interesting that this algebraic approach matches up to $\log\log n$ factors the running time bound of the purely
combinatorial algorithm of Bencz{\'u}r and Karger \cite{BenczurK96} for the computation of the (much weaker)
cut-preserving sparsifier. We note however that an $\tilde{O}(m+n\log^4 n)$ time cut-preserving sparsification algorithm was recently announced informally  \cite{HariharanP10}.

Sparsifying once with the Spielman and Srivastava algorithm and then applying our incremental sparsifier gives a $(k,O(k \log^3 n))$ ultrasparsifier that runs in $\tilde{O}(m \log^3 n)$  randomized time. Within the envelope of nearly-linear
time algorithms, this becomes the best known ultrasparsification algorithm in terms of both its quality and its running
time. Our guarantee on the quality of the ultrasparsifier is off by a factor of $O(\log^2 n)$ comparing to the
ultrasparsifier presented in \cite{Kolla10}. In the special case where the input graph has $O(n)$ edges, our incremental sparsifier is a $(k,O(k \log^2 n))$ ultrasparsifier.

%% file: sample.tex
\section{Sparsification by Oversampling} \label{sec:sample}

In this section we revisit a sampling scheme proposed by Spielman and Srivastava for sparsifying a graph \cite{SpielmanSrivastava08}. Consider the following general sampling scheme:

\begin{algo}[h]
\qquad

\textsc{Sample}
\vspace{0.05cm}

\underline{Input:} Graph $G=(V,E,w)$, $p':E
\rightarrow \R+$, real $\xi$.

\underline{Output:} Graph $G'=(V,E',w')$.
\vspace{0.2cm}

\begin{algorithmic}[1]
\STATE{$t:= \sum_e p'_e$}
\STATE{$q:= C_s  t \log{t} \log(1/\xi)$} {\footnotesize ~~~(* $C_S$ is a known constant  *)}
\STATE{$p_e:= {p'_e}/{t}$}
\STATE{$G':= (V,E',w')$ with $E'=\emptyset$}
\FOR{$q$ \mbox{times}}
\STATE{{\small Sample one $e \in E$ with probability of picking $e$ being $p_e$.} }
\STATE{{\small Add $e$ to $E'$ with weight $w'_e = w_e/p_e$}}
\ENDFOR
\STATE{For all $e\in E'$, let $w_{e'}: = w_e/q$}
\RETURN{$G'$}
\end{algorithmic}
\end{algo}

Spielman and Srivastava pick $p'_e=w_eR_e$ where $R_e$ is the effective resistance of $e$ in $G$, if $G$ is viewed as an electrical network with resistances $1/w_e$. This choice returns a spectral sparsifier. A key to bounding the number of required samples is the identity $\sum_{e} p_e'=n-1$. Calculating good approximations to the effective resistances seems to be at least as hard as solving a system, but as we will see in Section \ref{sec:incrementalsparsify}, it is easier to compute numbers $p_e'\geq (w_eR_e)$, while still controlling the size of $t=\sum_{e} p_e'$. The following Theorem considers a sampling scheme based on $p_e'$'s with this property.

\begin{theorem}\label{thm:sample}

\textbf{(Oversampling)} Let $G=(V,E,w)$ be a graph. Assuming that $p'_e \geq w_eR_e$ for each edge $e\in E$, and $\xi \in \Omega(1/n)$, the graph $G' = \textsc{Sample}(G, p', \xi)$ satisfies
$$
       G \preceq 2G' \preceq 3G
$$
with probability at least $1 - \xi $.
\end{theorem}

The proof follows closely that Spielman and Srivastava \cite{SpielmanSrivastava08}, with only a minor difference in one calculation. Let us first review some necessary lemmas.

If we assign arbitrary orientations on the edges, then we can define the incidence matrix $\Gamma \in \Re^{m \times n}$ as follows:

$$\Gamma_{e, u} =
\left\{
\begin{array}{lr}
-1 & \text{if u is the head of e} \\
1 & \text{if u is the tail of e} \\
0 & \text{otherwise}
\end{array}
\right.$$

If we let $W$ be the diagonal matrix containing edge weights, then $W^{1/2}$ is a real positive diagonal matrix as well since all edge weights are positive. The Laplacian $L$ can be written in terms of $W$ and $\Gamma$ as
$$L = \Gamma^T W \Gamma = \Gamma^T W^{1/2} W^{1/2} \Gamma. $$

Algorithm \textsc{Sample} forms a new graph by multiplying each edge $e$ by a nonnegative number $s_e$. If ${\mathbf S}$ is the diagonal matrix with $S(e,e) = s_e$, the Laplacian of the new graph can be seen to be equal to
$$\tilde{L} = \Gamma^T W \Gamma = \Gamma^T W^{1/2}{\mathbf S} W^{1/2} \Gamma. $$

Let $L^+$ denote the Moore-Penrose of $L$, i.e. the unique matrix sharing with $L$ its null space, and acting as the inverse of $L$ in its range. The key to the proofs of \cite{SpielmanSrivastava08} is the $m\times m$ matrix
$$
   \Pi = W^{1/2} \Gamma L^{+} \Gamma^T W^{1/2},
$$
for which the following lemmas are proved.

\begin{lemma} (Lemma 3i in \cite{SpielmanSrivastava08})
$\Pi$ is a projection matrix, i.e. $\Pi^2 = \Pi$.
\end{lemma}

\begin{lemma} (Lemma 4 in \cite{SpielmanSrivastava08}) \label{th:PiLap}
$$(1-||\Pi\Pi - \Pi S \Pi||_2)L \preceq \tilde{L} \preceq (1+||\Pi\Pi - \Pi S \Pi||_2)L.$$
\end{lemma}

We also use Lemma~\ref{thm:RudelsonVershynin} below, which is Theorem 3.1 from Rudelson and Vershynin \cite{RudelsonVershynin07}. The first part of the Lemma was also used as Lemma 5 in \cite{SpielmanSrivastava08} in a similar way.

\begin{lemma}\label{thm:RudelsonVershynin}
Let $p$ be a probability distribution over $\Omega \subseteq R^d$ such that $\sup_{y \in \Omega}||y||_2 \leq M$ and
$ || {\mathbb E}_p(yy^T) ||_2 \leq 1$. Let $y_1 \dots y_q$ be independent samples drawn from $p$, and let
$$a = CM\sqrt{\frac{\log{q}}{q}}.$$
Then:
\begin{enumerate}

\item $${\mathbb E}||\frac{1}{q} \sum_{i=1}^q y_iy_i^T - {\mathbb E}(yy^T)||_2 \leq a.$$

\item $$Pr [ ||\frac{1}{q} \sum_{i=1}^q y_iy_i^T - {\mathbb E}(yy^T)||_2 > x ] \leq {2} {e^{-cx^2/a^2}}.$$

\end{enumerate}
Here $C$ and $c$ are fixed constants.
\end{lemma}

\Proof (of Theorem~\ref{thm:sample})
 Following the pseudocode of \textsc{Sample}, let $t = \sum_e p_e'$ and $q=C_s t\log t \log (1/\xi)$. It can be seen that
 $$
     \Pi S \Pi = \frac{1}{q} \sum_{i=1}^q y_iy_i^T,
 $$
 where the $y_i$  are drawn from the distribution
$$
    y = \frac{1}{\sqrt{p_e}}\Pi(\cdot,e) \textnormal{~~with probability $p_e$}.
$$
For the distribution $y$ we have $E(yy^T) = \Pi\Pi = \Pi$. Since $\Pi$ is a projection matrix, we have $||\Pi||_2 \leq 1$. So, the condition imposed by Lemma \ref{thm:RudelsonVershynin} on the distribution holds for $y$.  The fact that $\Pi$ is a projection matrix also gives $$\Pi(:, e)^T\Pi(:, e) = (\Pi\Pi)(e,e) = \Pi(e,e),$$ which we use to bound $M$ as follows.
\begin{equation} \label{eq:upperboundM}
M =  \sup_{e} \frac{1}{\sqrt{p_e}} ||\Pi(:, e)||_2 =
 \sup_{e} \frac{1}{\sqrt{p_e}} \sqrt{\Pi(e, e)}  = 
   \sup_e \sqrt{\frac{t}{p'_e}} \sqrt{ w_eR_e } \leq \sqrt{t}.
\end{equation}

The last inequality follows from the assumption about the $p_e'$. Recall now that we have $\log(1/\xi) \leq \log{n}$ by assumption, $t \geq \sum_e w_eR_e$ by construction, and $\sum_e w_eR_e=n-1$ by Lemma 3 in \cite{SpielmanSrivastava08}. Combining these facts and setting $q = c_St \log t \log (1/\xi)$
for a proper constant $c_S$, part 1 of  Lemma \ref{thm:RudelsonVershynin} gives
$$
    a \leq \sqrt{\frac{4}{c\log(2/\xi)}}.
$$
Now substituting $x = \frac{1}{2}$ into part 2 of Lemma \ref{thm:RudelsonVershynin}, we get
\begin{eqnarray*}
   Pr [ ||\frac{1}{q} \sum_{i=1}^q y_iy_i^T - E(yy^T)||_2 > 1/2 ] \leq  
 {2} {e^{-(c/4)/a^2}}
\leq 2 {e^{(-c/4)/(4/c\log{2/\xi})}}
\leq \xi.
\end{eqnarray*}
It follows that with probability at least $1-\xi$ we have
 $$
     ||\frac{1}{q} \sum_{i=1}^q y_iy_i^T - E(yy^T)||_2 \leq 1/2,
 $$
 which implies $||\Pi S \Pi - \Pi\Pi||_2 \leq 1/2$. The theorem then follows by Lemma \ref{th:PiLap}.
\QED

{\bf Note.} The upper bound for $M$ in inequality \ref{eq:upperboundM} is in fact the only place where our proof differs from that of \cite{SpielmanSrivastava08}. In their case the last inequality is replaced by an exact inequality, which is possible because the exact values for $w_eR_e$ are used. In our case, by using inexact values we get a weaker upper bound which reflects in the density (depending on $m$, not $n$) of the incremental sparsifier. It is however enough for the solver.

%% file: incrementalsparsify.tex
\section{Incremental Sparsifier} \label{sec:incrementalsparsify}

Consider a spanning tree $T$ of $G = (V,E,w)$.  Let $w'(e) = 1/{w(e)}$. If the unique path connecting the endpoints of $e$ consists of edges $e_1 \dots e_k$, the stretch of $e$ by $T$ is defined to be
$$stretch_T(e) = \frac{\sum_{i=1}^k w'(e_i)}{w'(e)}.$$
Let $R_e$ denote the effective resistance of $e$ in $G$ and $RT_e$ denote the effective resistance of $e$ in $T$. We have $RT_e = \sum_{i=1}^k 1/w(e_i) $. Thus $stretch_T(e)  = w_e RT_e$.
By Rayleigh's monotonicity law \cite{doyle-2000}, we have $RT_e \geq R_e$, so $stretch_T(e) \geq w_eR_e$. As the numbers $stretch_T(e)$ satisfy the condition of Theorem \ref{thm:sample}, we can use them for oversampling. But at the same time we want to control the total stretch, as it will directly affect the total number of samples required in \textsc{SAMPLE}. This leads to taking $T$ to be a {\em low-stretch tree}, with the guarantees provided by the following result of Abraham, Bartal, and Neiman \cite{AbrahamBN08}.
\begin{theorem} (Corollary 6 in \cite{AbrahamBN08})
Given a graph $G=(V, E, w')$, \textsc{LowStretchTree(G)} in time $O(m \log n + n \log^2n)$, outputs a
spanning tree $T$ of $G$ satisfying
$\sum_{e \in E} =  O(m \log{n} \cdot \log{\log{n}}^3).$
\end{theorem}

Our key idea is to scale up the low-stretch tree by a factor of $\kappa$, incurring a condition number of $\kappa$ but allowing us to sample the non-tree edges aggressively using the upper bounds on their effective resistances given by the tree. The details are given in algorithm \textsc{IncrementalSparsify}.

\begin{algo}[h]
\qquad

\textsc{IncrementalSparsify}
\vspace{0.05cm}

\underline{Input:} Graph $G$, reals $\kappa$, $0<\xi<1$

\underline{Output:} Graph $H$
\vspace{0.2cm}

\begin{algorithmic}[1]
\STATE{$T := $ \textsc{LowStretchTree}(G)}
\STATE{Let $T'$ be $T$ scaled by a factor of $\kappa$}
\STATE{Let $G'$ be the graph obtained from $G$ \\ by replacing $T$ by $T'$}
\FOR{$e \in E$}
\STATE{Calculate $stretch_{T'}(e)$}
\ENDFOR
\STATE{$H := $ \textsc{Sample}($G'$, $stretch_{T'}$, $1/2 \xi$)}
\RETURN{$2H$}
\end{algorithmic}
\end{algo}

\begin{theorem} \label{th:incrementalsparsify} Given a graph $G$ with
  $n$ vertices, $m$ edges and any values $\kappa<m$, $\xi \in
  \Omega(1/n)$, \textsc{IncrementalSparsify} computes a graph $H$
such that:
\begin{itemize}
\item
$G  \preceq H \preceq 3 \kappa G$
\item
$H$ has  $n - 1+  \tilde{O}((m/\kappa) \log^2{n} \log(1/\xi) )$ edges,
\end{itemize}
with probability at least $1 - \xi$. The algorithm runs
in $\tilde{O} (m\log n +(n\log^2{n} + m \log^3 n /\kappa ) \log(1/\xi) )$ time.
\end{theorem}

\Proof We first bound the condition number. Since the weight of an edge is increased by at most a factor of $\kappa$, we have $G \preceq G' \preceq \kappa G$.
Furthermore, the effective resistance along the tree of each non-tree edge decreases by a factor of $\kappa$.
Thus \textsc{IncrementalSparsify} sets $p'_e = 1$ if $e \in T$ and $stretch_T(e)/\kappa$ otherwise, and invokes \textsc{Sample} to compute a graph $H$ such that with probability at least $1-\xi$, we get
$$G\preceq G' \preceq H \preceq {3}G' \preceq 3\kappa G.$$

We next bound the number of non-tree edges. Let $t' = \sum_{e \notin   T}stretch_{T'}(e) $, so $t' = \tilde{O}((m/\kappa) \log n)$. Then for the number $t$ used in \textsc{Sample} we have $t = t'+n-1$ and $q
= C_s t\log t \log(1/\xi)$ is the number of edges sampled in the call of \textsc{Sample}. Let $X_i$ be a random variable which is $1$ if the $i^{th}$ edge picked by \textsc{Sample} is a non-tree edge and
$0$ otherwise. The total number of non-tree edges sampled is the
random variable $X = \sum_{i=1}^q X_i$, and its expected value can be
calculated using the fact $Pr(X_i=1)=t'/t$:
\begin{eqnarray*}
   E[X] & =  &q \frac{t'}{t} = t' \frac{C_s t\log t \log(1/\xi) }{\kappa t} =  \tilde{O}((m/\kappa)\log^2 n \log(1/\xi)).
\end{eqnarray*}
A standard form of Chernoff's inequality is:
$$
    Pr[X>(1+\delta)E[X]] <\left( \frac{e^\delta}{(1+\delta)^{(1+\delta)}} \right)^{E[X]}.
$$
Letting $\delta =2$, and using the assumption $k<m$, we get
    $Pr(X>3E[X])< ( e^2/27)^{E[X]} < 1/n^c,
$
for any constant $c$. Hence, the probability that \textsc{IncrementalSparsify} succeeds, with respect to both the number of non-tree edges and the condition number, is at least $1-\xi$.

We now consider the time complexity. We first generate a low-stretch spanning tree in $O(m \log n + n \log^2 n)$ time.  We then compute the effective resistance of each non-tree edge by the tree.
This can be done using Tarjan's off-line LCA algorithm \cite{Tarjan79}, which takes $O(m)$ time \cite{GabowTarjan83}.
We next call \textsc{SAMPLE} with parameters that make it draw $\tilde{O}((n+m/\kappa \log n)\log n \log (1/\xi))$ samples (precisely, $O(t \log t \log (1/\xi))$ samples where $t = \tilde{O}(n+m/\kappa \log n)$). To compute each sample efficiently, we assign each edge an interval on the unit interval $[0,1]$ with length corresponding to its probability, so that no two intervals overlap. At each sampling iteration we pick a random value in $[0,1]$ and do a binary search in order to find the interval that contains it in $O(\log{n})$ time. Thus the cost of a call to \textsc{SAMPLE} is $\tilde{O}((n \log^2 n +m/\kappa \log^3 n )\log (1/\xi))$. \QED

%% file: solver.tex
\section{Solving using Incremental Sparsifiers} \label{sec:solver}

The solver of Spielman and Teng \cite{SpielmanTeng08c} consists of two phases. The {\em preconditioning phase} builds a chain of progressively smaller graphs ${\cal C} = \{A_1,B_1,A_2,\ldots,A_d\}$ starting with $A_1=A$. The process for building ${\cal C}$ alternates between calls to a sparsification routine \textsc{UltraSparsify} which constructs $B_i$ from $A_i$ and a routine \textsc{GreedyElimination} (following below) which constructs $A_{i+1}$ from $B_i$. The preconditioning phase is independent from the $b$-side of the system $L_Ax=b$.

\begin{algo}[h]
\qquad

\textsc{GreedyElimination}
\vspace{0.05cm}

\underline{Input:} Weighted graph $G=(V,E,w)$\\
\underline{Output:} Weighted graph $\hat{G}=(\hat{V},\hat{E},\hat{w})$
\vspace{0.2cm}

\begin{algorithmic}[1]
\STATE{$\hat{G}:=G$}
\REPEAT
\STATE{greedily remove all degree-$1$ nodes from $\hat{G}$}
\IF{$deg_{\hat{G}}(v)=2$ and $(v,u_1),(v,u_2)\in E_{\hat{G}}$}
\STATE{$ w' := w(u_1,v)w(u_2,v)/\left(w(u_1,v)+w(u_2,v)\right)$}
\STATE{replace $(u_1,v,u_2)$ by an edge of weight $w'$ in $\hat{G}$}
\ENDIF
\UNTIL{there are no nodes of degree $1$ or $2$ in $\hat{G}$ }
\RETURN {$\hat{G}$}
\end{algorithmic}
\end{algo}

The {\em solve phase} passes $\cal C$, $b$ and a number of iterations $t$ (depending on a desired error $\epsilon$) to the recursive preconditioning algorithm \textsc{R-P-Chebyshev}, described in Section \ref{sec:rpcheb}. The time complexity of the solve phase depends on $\epsilon$, but more crucially on the quality of $\cal C$, which is a function of the sparsifier quality. 

\begin{definition} [\textbf{$\kappa(n)$-good chain}]
Let $\kappa(n)$ be a monotonically non-decreasing function of $n$. Let ${\cal C} = \{A = A_1,B_1,A_2,\ldots,A_d\}$ be a chain of graphs, and denote by $n_i$ and $m_i$ the numbers of nodes and edges of $A_i$ respectively. We say that $\cal C$ is  $\kappa(n)$-good for $A$, if:
  \begin{enumerate}
   \item $A_i \preceq B_i \preceq \kappa(n_i) A_i$.
   \item $A_{i+1} = \textsc{GreedyElimination}(B_i)$.
   \item $m_i/m_{i+1} \geq c_r \sqrt{\kappa(n_i)}$, {for some constant $c_r$.}
  \end{enumerate}
\end{definition}

Spielman and Teng analyzed a recursive preconditioned Chebyshev iteration and showed that a $\kappa(n)$-good chain for $A$ can be used to solve a system on $L_A$. This is captured by the following Lemma, adapted from Theorem 5.5 in \cite{SpielmanTeng08c}.
\vspace{0.2cm}
\begin{lemma} \label{th:requirement}
  Given a $\kappa(n)$-good chain for $A$, a vector ${x}$ such that $||{x}-L_A^{+}b||_A<\epsilon ||L_A^{+}b||_A$ can be computed in ${O}(m_d^3m_1 \sqrt{\kappa(n_1)}  \log(1/\epsilon))$ expected time.
\end{lemma}

\vspace{0.2cm}
 For our solver, we follow the approach of Spielman and Teng. The main difference is that we replace their routine \textsc{UltraSparsify} with our routine \textsc{IncrementalSparsify}, which is not only faster but also constructs a better chain which translates into a faster solve phase. We are now ready to state our algorithm for building the chain. In what follows we write $v:=O(g(n_i))$ to mean `$v:=f(n_i)$ for some explicitly known function $f(n) \in O(g(n))$'.

\begin{algo}[h]
\qquad

\textsc{BuildChain}
\vspace{0.05cm}

\underline{Input:} Graph $A$, scalar $p$ with $0<p<1$ \\
\underline{Output:} Chain of graphs ${\cal C}= \{A=A_1,B_1,A_2,\ldots,A_d\}$
\vspace{0.2cm}

\begin{algorithmic}[1]
 \STATE $A_1 := A$
 \STATE ${\cal C}:=\emptyset$ 
 \WHILE{$m_i > (\log \log n)^{1/3}$}
 \IF{$m_i>\log n$}
 \STATE $\xi:=\log n$
 \ELSE
 \STATE $\xi: = \log \log n$
 \ENDIF
 \STATE $\kappa := \tilde{O}(\log^4 n_i \log(1/p))$
 \STATE $B_i := \textsc{IncrementalSparsify}(A_i, \kappa ,p/(2\xi) )$
 \STATE  $A_{i+1} := \textsc{GreedyElimination}(B_i)$
 \IF{$m_i/m_{i+1} <  c_r \sqrt{3\kappa}$}
 \RETURN {FAILURE}
 \ENDIF
 \STATE ${\cal C} = {\cal C}\cup\{A_i,B_i\}$
 \STATE $i: = i+1$
 \ENDWHILE
 \RETURN {$\cal C$}
\end{algorithmic}
\end{algo}

\begin{lemma} \label{th:incrementalok}
 Given a graph $A$, $\textsc{BuildChain(A,p)}$ produces an $\tilde{O}(\log^4 n)$-good chain for $A$, with probability at least $1-p$. The algorithm runs in time $$\tilde{O}((m\log n +n \log^2 n)\log(1/p)).$$
\end{lemma}
\Proof Assume that $B_i$ has $n_i-1+m_i/k'$ edges. A key property of $\textsc{GreedyElimination}$ is that if $G$ is a graph with $n-1+j$ edges, \textsc{GreedyElimination}$(G)$ has at most  $2j-2$ vertices and $3j-3$ edges \cite{SpielmanTeng08c}. Hence
$\textsc{GreedyElimination}(B_i)$ has at most $3m_i/k'$ edges. It follows that $m_i/m_{i+1} \geq k'/3$. Then, in order to satisfy the second requirement, we must have $A_i\preceq  B_i \preceq c' k'^2 A_i$, for some sufficiently small constant $c'$.


However, we also know that the call to \textsc{IncrementalSparsify} returns an incremental sparsifier $B_i$ that $3\kappa$-approximates $A_i$. So it is necessary that $c'k'^2 > 3\kappa$. Moreover, $B_i$ has $n_i-1+\tilde{O}(m_i\log^2 n/\kappa)$ edges, a number  which we assumed is equal to $n_i-1+m_i/k'$. The value assigned to $\kappa$ by the algorithm is taken to be the minimum that satisfies these two conditions.

The probability that $B_i$ has the above properties is by construction at least $1-p/(2\log n)$ if $n_i>\log n$ and $1-p/(2\log \log n)$ otherwise. The probability that the requirements hold for all $i$ is then at least
\begin{eqnarray*}
(1-p/(2\log n))^{\log n}(1-p/(2\log \log n))^{\log \log n} \\ >(1-p/2)^2>1-p.
\end{eqnarray*}

Finally note that each call to \textsc{IncrementalSparsify} takes $\tilde{O}((m_i \log^2 n)\log(1/p))$ time. Since $m_i$ decreases faster than geometrically with $i$, the claim about the running time follows. \QED


Combining Lemmas \ref{th:requirement} and \ref{th:incrementalok} proves our main Theorem.

\begin{theorem}
 On input an $n\times n$ symmetric diagonally dominant matrix $A$ with $m$ non-zero entries and a vector $b$,  a vector ${x}$ satisfying $||{x}-A^{+}b||_A<\epsilon ||A^{+}b||_A $, can be computed in
 expected time $\tilde{O}(m\log^2{n}\log(1/\epsilon)).$
\end{theorem}

%% file: extensions.tex
\section{Comments / Extensions} \label{sec:logsample}

Unraveling the analysis of our bound for the condition number of the incremental sparsifier, it can been that one $\log n$ factor is due to the number of samples required by the Rudelson and Vershynin theorem. The second $\log n$ factor is due to the average stretch of the low-stretch tree.

It is quite possible that the low-stretch construction and perhaps the associated lower bound can be bypassed -at least for some graphs- by a simpler approach similar to that of \cite{KoutisMi07}. Consider for example the case of unweighted graphs. With a simple ball-growing procedure we can concede in our incremental sparsifier a $1/\log n$ fraction of the edges, while keeping within clusters of diameters $O(\log^2 n)$ the rest of the edges. The design of low-stretch trees may be simplified within the small diameter clusters. This diameter-restricted local sparsification is a natural idea to pursue, at least in an actual implementation of the algorithm.

%% file: Paper.bbl
\newcommand{\etalchar}[1]{$^{#1}$}

%% file: rpcheb.tex
\section{Appendix: The Complete Solver} \label{sec:rpcheb}

The purpose of this section is to provide a few more algebraic details about the chain of preconditioners, and the recursive preconditioned Chebyshev method which consists the solve phase of the solver. The material is not new and we include it only for completeness. We focus on pseudocode. We refer the reader to \cite{SpielmanTeng08c} for a more detailed exposition along with proofs.

{\bf Direct methods - Cholesky factorization.} If $A$ is a symmetric and positive definite (SPD) matrix, it can be written in the form $A = LL^T$, a product known as the {\em Cholesky factorization} of $A$. This extends to Laplacians, with some care for the null space. The Cholesky factorization can be computed via a symmetric version of {\em Gaussian elimination}. Given the decomposition, solving the systems $L y =  b$ and $ L^T x = y$ yields the solution to the system $Ax=b$; the key here is that solving with $L$ and $L^T$ can be done easily via forward and back substitution. A {\em partial Cholesky factorization} with respect to the first $k$ variables of $A$, puts it into the form
\begin{equation} \label{eq:partialCholesky}
     A = L \left(
             \begin{array}{cc}
               I_k & 0 \\
               0 & A_k \\
             \end{array}
           \right) L^T
\end{equation}
where $I_k$ denotes the $k\times k$ identity matrix, and $A_k$ is known as the Schur complement of $A$ with respect to the elimination of the $k$ first variables. The matrix $A_{k+1}$ is the Schur complement of $A_k$ with respect the the elimination of its first variable.

Given a matrix $A$, the graph $G_A$ of $A$ is defined by identifying the vertices of $G_A$ with the rows and columns of $A$ and letting the edges of $G_A$ encode the non-zero structure of $A$ in the obvious way.

It is instructive to take a graph-theoretic look at the partial Cholesky factorization when $k=1$. In this case, the graph $G_{A_1}$ contains a clique on the neighbors of the first node in $G_A$. In addition, the first column of $L$ is non-zero on the corresponding coordinates. This problem is known as {\em fill}. It then becomes obvious that the complexity of computing the Cholesky factorization depends crucially on the ordering of $A$. Roughly speaking, a good ordering has the property that the degrees of the top nodes of $A,A_1,A_2,\ldots,A_k$ are as small as possible. The best known algorithm for positive definite systems of planar structure runs in time $O(n^{1.5})$ and it is based on the computation of good orderings via nested dissection \cite{george73nested,lipton79generalized, AlonYuster10}.

There are two fairly simple but important facts considering the partial Cholesky factorization of equality \ref{eq:partialCholesky} \cite{SpielmanTeng08c}.
First, if the top nodes of $A,A_1,\ldots,A_{k-1}$ have degrees $1$ or $2$, then back-substitution with $L$ requires only $O(n)$ time. Second, if $A$ is a Laplacian, then $A_k$ is a Laplacian. Such an ordering and the corresponding Laplacian $A_k$ can be found in linear time via \textsc{GreedyElimination}, described in Section \ref{sec:solver}. The corresponding factor $L$ can also be computed  easily.

{\bf Iterative methods.} Unless the system matrix is very special, direct methods do not yield nearly-linear time algorithms. For example, the nested dissection algorithm is known to be asymptotically optimal for the class of planar SPD systems, within the envelope of direct methods. Iterative methods work around the fill problem by producing a sequence of approximate solutions using only matrix-vector multiplications and simple vector-vector operations. For example {\em Richardson's iteration} generates an approximate solution $x_{i+1}$ from $x_i$, by letting
$$
   x_{i+1} = (I-A)x_i + b.
$$

The solver in this paper, as well as the Spielman and Teng solver \cite{SpielmanTeng08c}, are based on the very well studied Chebyshev iteration \cite{Axe94}. The {\em preconditioned Chebyshev iteration} (\textsc{P-Chebyshev}) is the Chebyshev iteration applied to the system $B^{+}A x = B^{+}b$, where $A,B$ are SPD matrices, and $B$ is known as the {\em preconditioner}. The preconditioner $B$ needs not be explicitly known. The iteration requires matrix-vector products with $A$ and $B^{+}$. A product of the form $B^{+1}z$ is equivalent to solving the system $B y = c$. Therefore (\textsc{P-Chebyshev}) requires access to only a function $f_B(c)$ returning $B^{+1}c$. In addition it requires a lower bound $\lambda_{\min}$ on the minimum eigenvalue of $(A,B)$ and an upper bound $\lambda_{\max}$ on the maximum generalized eigenvalue of $(A,B)$.


\begin{algo}[h]

\vspace{0.05cm}
\textsc{P-Chebyshev}

\underline{Input:} SPD matrix $A$, vector $b$, number of iterations $t$,
\\ preconditioner $f_B(z)$, $\lambda_{min}$, $\lambda_{max}$

\underline{Output:} approximate solution $x$ for $Ax=b$
\vspace{0.2cm}

\begin{algorithmic}
 \STATE $x:= 0$
 \STATE $r:= b$
 \STATE $d := (\lambda_{max} + \lambda_{min})/2$
 \STATE $c := (\lambda_{max} - \lambda_{min})/2$
 \FOR {$i=1 \mbox{ to } t$}
 \STATE $z := f_B(r)$
 \IF {$i=1$}
 \STATE $x:=z$
 \STATE $\alpha := 2/d$
 \ELSE
 \STATE $\beta := (c \alpha / 2 )^2$
 \STATE $\alpha := 1/(d - \beta)$
 \STATE $x := z + \beta x$
 \ENDIF
 \STATE $x  := x + \alpha x$
 \STATE $r  := b - Ax$
 \ENDFOR
 \RETURN{$x$}\\
\end{algorithmic}
\end{algo}
A well known fact about the Chebyshev method is that after  $O(\sqrt{\lambda_{\max}/\lambda_{\min}} \log 1/\epsilon)$ iterations the return vector $x$ satisfies $\norm{{x}-A^+ b}_A \leq \epsilon \norm{A^+b}_A$ \cite{Axe94}.

\vspace{0.2cm}
{\bf Hybrid methods.} One of the key ideas in Vaidya's approach was to combine direct and iterative methods into a hybrid method by exploiting properties of Laplacians. \cite{Vaidya91}. For the rest of this section we will identify graphs and their Laplacians, using their natural 1-1 correspondence.

Let $A_1$ be a Laplacian. The incremental sparsifier $B_1$ of $A_1$ is a natural choice as preconditioner. With proper input parameters, \textsc{IncrementalSparsify} returns a $B_1$ that contains enough degree $1$ and $2$ nodes, so that \textsc{GreedyElimination} can make enough progress reducing $B_1$ to a matrix of the form
$$
     B_1 = L_1 \left(
             \begin{array}{cc}
               I & 0 \\
               0 & A_2 \\
             \end{array}
           \right) L_1^T,
$$
where $A_2$ is the output of algorithm \textsc{GreedyElimination}. Let $I_j$ denote the identity of dimension $j$ and
\begin{eqnarray*}
   \Pi_1 & = & \left(
            \begin{array}{cc}
              0 & I_{dim(A_2)} \\
            \end{array}
          \right) \\
       Q_1 & = & \left(
            \begin{array}{cc}
             I_{dim(A_1)-dim(A_2)} & 0 \\
            \end{array}
          \right).
\end{eqnarray*}
Recall that \textsc{P-Chebyshev} requires the solution of $By =c$, which is given by
$$
   y = L_1^{-T} \left(
                \begin{array}{c}
                  Q_1 L_1^{-1}c \\
                  A_1^{+} \Pi_1 L_1^{-1}c  \\
                \end{array}
              \right).
$$
The two matrix-vector products with $L_1^{-1},L_1^{-T}$ can be computed in time $O(n)$ via forward and back substitution. Therefore, we can solve a system in $B$ by solving a linear system in $A_2$ and performing $O(n)$ additional work. Naturally, in order to solve systems on $A_2$ we can recursively apply preconditioned Chebyshev iterations on it, with a new preconditioner $B_2$. This defines a preconditioning chain $\cal C$ that consists of progressively smaller graphs $A=A_1,B_1,A_2,B_2,\ldots,A_d$, along with the corresponding matrices $L_i,\Pi_i,Q_i$ for $1\leq i\leq d-1$. So, to be more precise than in Section \ref{sec:solver}, routine \textsc{BuildChain} has the following specifications.

\begin{algo}[h]
\qquad

\textsc{BuildChain}
\vspace{0.05cm}

\underline{Input:} Graph $A$, scalar $p$ with $0<p<1$ \\
\underline{Output:} Chain ${\cal C} = \{ \{A_i,B_i,L_i,\Pi_i,Q_i\}_{i-1}^{d-1},A_d \}$
\vspace{0.2cm}

\end{algo}

We are now ready to give pseudocode for the recursive preconditioned Chebyshev iteration.
\begin{algo}[h]

\qquad

{\textsc{R-P-Chebyshev}}

\underline{Input:} Chain $\cal C$, level $i$, vector $b$, number of iterations $t$\\
\underline{Output:} Approximate solution ${x}$ for $A_ix = b$

 \vspace{0.2cm}
\begin{algorithmic}[1]
 \IF {$i=d$ for some fixed $d$}
 \RETURN{$A_i^{+} b$}
 \ELSE
 \STATE $\kappa := \kappa(A_i,B_i)$
 \STATE Define function $f_i(z)$:
 \STATE \qquad $t' := \lceil 1.33 \sqrt{\kappa} \rceil$
 \STATE \qquad $z' := L_i^{-1} z$
 \STATE \qquad $z_1'' := Q_{i} z'$
 \STATE \qquad $z_2'' := \textsc{R-P-Chebyshev}({\cal C},i+1, \Pi_{i}z',t')$
 \STATE \qquad $f_i(z) \gets L_i^{-T} [z_1''~~ z_2'']^T$
 \STATE $l : = 1-2e^{-2}$
 \STATE $u : = (1+2e^{-2})\kappa$
 \STATE ${x}:=$\textsc{P-Chebyshev}$(A_i,b,t,f_i(z),l,u)$
 \RETURN {x}
 \ENDIF
\end{algorithmic}
\end{algo}

{\bf The complete solver.} Finally, the pseudocode for the complete solver is as follows.


\begin{algo}[h]

\qquad

{\textsc{Solve}}

\underline{Input:} Laplacian $L_A$, vector $b$,
error $\epsilon$, failure probability $p$ \\
\underline{Output:} Approximate solution ${x}$

 \vspace{0.2cm}
\begin{algorithmic}
\STATE ${\cal C} := \textsc{BuildChain}(A,p)$
\STATE $x := \textsc{R-P-Chebyshev}({\cal C},1,b,\tilde{O}(\log^2 n \log (1/\epsilon))$
\end{algorithmic}
\end{algo}